\documentclass[useAMS,usenatbib]{mn2e}
\usepackage{graphicx}
\usepackage{times,epsfig}

\newcommand{\ha}{H$\alpha$} 
\newcommand{\hbeta}{H$\beta$}
\newcommand{\helium}{He{\sc i}}
\newcommand{\heliumb}{He{\sc ii}}

\newcommand{\oi}{[O~{\sc i}]} 
 
\newcommand{\NII}{[N~{\sc ii}]~6548~\&~6584~\AA}

\newcommand{\oiii}{[O~{\sc iii}]~5007~\AA}

\newcommand{\oii}{[O~{\sc ii}]} 

\newcommand{\nitrogen}{[N~{\sc ii}]}
\newcommand{\nitrogena}{[N~{\sc i}]}
\newcommand{\oxygen}{[O~{\sc iii}]}
\newcommand{\sulfur}{[S~{\sc iii}]}
\newcommand{\sulfurt}{[S~{\sc ii}]}
\newcommand{\argonV}{[Ar~{\sc v}]}
\def\vhel{\ifmmode{V_{{\rm HEL}}}\else{$V_{{\rm HEL}}$}\fi}
\def\vsys{\ifmmode{V_{\rm sys}}\else{$V_{\rm sys}$}\fi}
\def\kms{\ifmmode{~{\rm km\,s}^{-1}}\else{~km~s$^{-1}$}\fi}
\def\vlsr{\ifmmode{v_{\rm lsr}}\else{$v_{\rm lsr}$}\fi}
\newcommand{\flux}{$10^{-16}$ erg s$^{-1}$ cm$^{-2}$ arcsec$^{-2}$}
\title[New PNe in the Galactic bulge -- II] {New Planetary Nebulae in
the Galactic bulge region with $l>0^{o}$~- II.}

\author[P. Boumis et al.] {P. Boumis$^{1}\thanks{e-mail:
ptb@astro.noa.gr}$, S. Akras$^{1,2}$, 
E. M. Xilouris$^{1}$, F. Mavromatakis$^{3}$, 
E. Kapakos$^{2}$, \newauthor J. Papamastorakis$^{2,4}$, and
C. D. Goudis$^{1,5}$\\
$^{1}$Institute of Astronomy \& Astrophysics,
National Observatory of Athens, I. Metaxa \& V. Paulou, GR--152 36
P. Penteli, Athens, Greece.\\
$^{2}$Department of Physics,University of Crete, P.O. Box 2208, GR-710 03 Heraklion, Crete,
Greece.\\
$^{3}$Technological Education Institute of Crete, General Department of Applied Science, P.O. Box 1939, GR-710 04 Heraklion, Crete, Greece.\\
$^{4}$Foundation for Research and Technology-Hellas,
P.O. Box 1527, GR-711 10 Heraklion, Crete, Greece.\\
$^{5}$Astronomical Laboratory, Department of Physics, University of Patras, GR-265 00
Rio-Patras, Greece.\\}

\begin{document}  

\date{Accepted 2006 January 9; Received 2005 December 28; in original form 2005 July 27}

\pagerange{\pageref{firstpage}--\pageref{lastpage}} \pubyear{2006}

\maketitle

\label{firstpage}

\begin{abstract}
 
\noindent The presentation of new results from an \oiii\ 
survey in a search for planetary nebulae (PNe) in the galactic
bulge is continued. 
A total of 60 objects, including 19 new PNe, have been detected 
in the remaining 34 per cent of the survey area, while 41 
objects are already known.
Deep \ha $+$\nitrogen~CCD images as well as low resolution spectra have been
acquired for these objects.
Their spectral signatures suggest that the detected emission
originates from photoionized nebulae. In addition, absolute line
fluxes have been measured and the electron densities are
given. Accurate optical positions and optical diameters are  also
determined.
\end{abstract}

\begin{keywords}
surveys -- planetary nebulae: general - Galaxy: bulge.
\end{keywords}

\section{Introduction}

This is the second of a series of papers presenting optical imaging and
spectrophotometric results on newly discovered planetary nebulae. 
In Paper I (\citealt{Bo03}), the discovery method and first results 
on the \oiii\ survey, in the northern Galactic bulge, were presented.

PNe are powerful tracers of our Galaxy's star formation history (Paper
I; \citealt{Be00} and references therein). The study of PNe can provide 
insight to the late stages of stellar
evolution, the nucleosynthesis in low and intermediate mass stars (1
M$_{\odot}$ to 8 M$_{\odot}$) and the chemical evolution of galaxies. When
a low or intermediate mass star is closing to its end, 
it goes through the red--giant (RG) phase, followed by a heavy
mass--loss period known as the asymptotic--giant--branch (AGB) stage.
Finally, it becomes a white--dwarf (WD) surrounded by a planetary nebula
(PN) made up of ionized gas from earlier stages of mass--loss. 

PNe near the Galactic center can be considered to lie, roughly, at the
same distance. Approximately 500 PNe have been discovered up to now, 
while $\sim$ 3500 have been discovered in our Galaxy (Paper I
and references therein; \citealt{Pa03,Pa05a,Pa05b};
\citealt{Ja04}). These numbers are considered small when compared with that
generally expected (15000 to 30000; \citealt{Ac92a}; \citealt{Zi91}).
Therefore, our Galaxy and especially, the Bulge area is well suited to
search for new PNe.

Information concerning the \oxygen~survey (observations and analysis)
are given in sect. 2, while in sect. 3 the follow-up
observations are presented (CCD imaging and spectroscopy) of the newly 
discovered PNe. 
In addition, flux measurements, diameter determination, accurate
positions and other physical properties for the new PNe are given is
sect. 4. In an forthcoming paper (Paper III -- in preparation), we
will present photoionization modelling applied to the new PNe, in
order to gain more insight into their physical parameters. 
In particular, flux calibrated images for most of the new PNe together 
with their spectroscopic results will be used in a PNe photoionization model
(CLOUDY; e.g. \citealt{Va99}).

\section{The \oiii\ Survey}

\subsection{Observations \& Results}

The observations were performed with the 0.3 m Schmidt-Cassegrain
(f/3.2) telescope at Skinakas Observatory in Crete, Greece in June
14--23 \& 26--28 and July 16--20, 2001. An
\oiii~interference filter with an 28 \AA~bandwidth was used in
combination with a Thomson CCD (1024$\times$1024). This configuration
results in a scale of 4.12 arcsec pixel$^{-1}$~and a field of view of
71 $\times$~71 arcmin$^{2}$~on the sky.  In our survey we observed
the regions $10^{o} < l < 20^{o}$, $-10^{o} < b < -3^{o}$ and $0^{o} <
l < 20^{o}$, $3^{o} < b < 10^{o}$ (Fields A and B in Fig. 1 - total
coverage on sky $\sim$220 square degrees).  The filled dark rectangles
in Fig. 1 represent the observed fields. The remaining
region of 34 per cent of the proposed grid (63 fields out of 179) is covered,
which was not observed because of the availability of the telescope.
All targets were observed between airmass 1.4 to 2.0 in similar observing
and seeing (1 -- 2 arcsec) conditions.

Two exposures in \oiii~of 1200 s and three exposures in the continuum,
each of 180 s, were obtained to identify and remove any cosmic ray hits. 
Two different continuum filters were used,
depending on their availability. Details about the filters, the
selection criteria and the detection method are given in Paper I.
After a detailed and systematic visual investigation of the
remaining 63 fields, 60 objects were identified. 
Images outlining the analysis procedure through the various steps 
can be found in Paper I.

After identifying the PNe candidates, a preliminary astrometry solution 
was obtained for all images containing one or more
candidates (see Paper I for details). 
Sixty (60) objects were detected in these images. In order to 
identify the already known PNe, we used up--to--date published catalogues 
related to planetary nebulae (see Table 3 in Paper I). The search showed that
among the objects found in our survey, 41 of them are known PNe, while 19
objects are new PNe candidates. Note that independent work by \cite{Pa03,Pa05b}
resulted in a new catalogue of PNe, where 9 of our new PNe candidates
are included as new or candidate PNe (see Table \ref{table1}).

\begin{table*}
\centering
\caption{Newly discovered PNe.}
\label{table1}
\begin{tabular}{c|c|c|c|c|c|c|c|c|c|c}
\hline
Object & PN G & RA & Dec & IRAS source  & F$_{12}$ & F$_{25}$ & F$_{60}$ & F$_{100}$ & F$_{\rm Quality}$ & Ref$^{\rm a}$ \\
 & (lll.l $\pm$ bb.b) & (J2000) & (J2000) & & (Jy) & (Jy) & (Jy) & (Jy) & (12,25,60,100 $\mu$m) & \\
\hline
PTB26 & 008.3$+$09.6 & 17 29 13.1 & $-$16 47 42.6 &  &  & &  & &  & 1b\\
PTB27 & 008.4$-$02.8 & 18 15 12.7 & $-$23 01 03.8 & & & & & & & 2a\\
PTB28 & 008.6$+$06.7 & 17 40 21.1 & $-$18 05 08.9 & 17374$-$1803 & 4.71 & 2.97 & 0.51 & 4.32 & 3,3,1,1 & \\
PTB29 & 008.7$-$04.2 & 18 21 08.2 & $-$23 23 56.9 & & & & & & & 2a\\
PTB30 & 010.1$+$04.4 & 17 51 46.6 & $-$18 04 05.0 & & & & & & & 1a\\
PTB31 & 011.0$-$02.9 & 18 20 53.7 & $-$20 48 10.9 & 18179$-$2049 & 0.36 & 0.67 & 3.1 & 88.6 & 1,3,3,1 & 1b\\
PTB32 & 011.3$-$09.1 & 18 45 10.2 & $-$23 21 39.7 & & & & & & & \\
PTB33 & 011.4$-$05.3 & 18 30 41.9 & $-$21 31 51.0 & & & & & & & 2b\\
PTB34 & 011.8$-$05.0 & 18 30 07.7 & $-$21 05 02.6 & & & & & & & \\
PTB35 & 012.1$-$02.6 & 18 21 43.7 & $-$19 39 45.7 & & & & & & & 1a \\
PTB36 & 013.2$-$05.0 & 18 32 45.3 & $-$19 49 32.2 & & & & & & & 1a \\
PTB37 & 013.7$-$04.7 & 18 32 34.6 & $-$19 14 03.7 & & & & & & & 1a\\
PTB38 & 013.8$-$02.0 & 18 23 04.3 & $-$17 53 31.5 & 18201$-$1755 & 1.58 & 1.05 & 22.4 & 258 & 1,3,1,1 & \\
PTB39 & 014.2$-$03.4 & 18 29 00.2 & $-$18 10 46.2 & & & & & & \\
PTB40 & 014.3$-$07.2 & 18 43 39.6 & $-$19 48 30.9 & & & & & & & \\
PTB41 & 014.8$-$02.7 & 18 27 26.6 & $-$17 24 10.0 & & & & & & & \\
PTB42 & 015.3$-$03.3 & 18 30 22.9 & $-$17 11 53.7 & & & & & & & \\
PTB43 & 016.6$-$04.0 & 18 35 55.8 & $-$16 21 20.5 & 18330$-$1623 & 1.14 & 1.46 & 4.4 & 68.8 & 3,3,1,1 & \\
PTB44 & 016.9$-$09.7 & 18 57 39.8 & $-$18 36 16.0 & & & & & & & \\
\hline
\end{tabular}

\medskip{}
\begin{flushleft}

${\rm ^a}$ Independently discovered by \cite{Pa03,Pa05b} as new PNe (1a, 2a) and candidate PNe (1b, 2b). \\ 

\end{flushleft}
\end{table*}

As in Paper I, a search in the IRAS Point Source Catalog (1988) was
performed for the presence of dust at the
positions of the new PNe candidates. The correlation revealed 4
matches (see Table~\ref{table1}). Taking into account their low flux
quality density, a reasonable number of them satisfy the standard
criteria (F$_{12} / {\rm F}_{25} \leq 0.35$~and F$_{25} / {\rm F}_{60}
\geq 0.3$ -- \citealt{Po88}; \citealt{Va95}) used to consider a new
object as a probable PN. A cross--check of our new PNe list was also
performed with the 2MASS Point Source Catalog (2000), the MSX infrared
astrometric Catalog \citep{Eg99} and the radio known PNe catalog. However, 
the results were negative and it may be that the low sensitivity of these 
surveys prevented a positive identification. 
In a very recent paper, \cite{Lu05} presented radio identifications for 
315 recently
discovered PNe using the 1400 MHz (NVSS) images. Since the radio
catalogue was just published, all our new PNe (including those of Paper I)
were examined. A detailed check of this catalogue showed
that there is a possibility of correlation between the radio sources and a
number of our new PNe. Following \cite{Lu05}, if the offset between
the optical and the radio position is less than 20\arcsec, the radio
source is probably identified with a PN, while offsets less than
10\arcsec are considered positive identifications. Therefore, 15 of our
new PNe were identified with radio sources. In Table~\ref{table2}, we
present the coordinates of the radio sources \citep{Lu05} which probably corellate
with our PNe, their measured radio flux and the difference in
arcseconds between the optical and the radio centres.

\begin{table*}
\centering
\caption{Radio sources \citep{Lu05} which probably are correlated with the new
PNe. Note that radio identification for the PNe of Paper I can also be
seen.}
\label{table2}
\begin{tabular}{c|c|c|c|c|c|c|c|c|c}
\hline
Object & RA$^{\rm a}$ & Dec$^{\rm a}$ & S$_{1.4\rm{GHz}}$ &  difference$^{\rm b}$ & \\
       & (J2000) & (J2000) & (mJy)  &    (arcsec) & \\
\hline
PTB01  & 17 43 38.88 & -24 31 58.2 & 4.3 & 6.81 & \\
PTB05  & 17 41 38.97 & -21 44 33.9 & 4.5 & 4.19 & \\
PTB11 & 18 13 40.63 & -23 57 38.0 & 2.7 & 5.93 &\\
PTB12 & 17 47 15.74 & -19 57 23.3 & 2.3 & 4.76 & \\
PTB13 & 17 51 07.74 & -19 25 40.8 & 3.0 & 14.8 & \\
PTB15 & 17 57 5.85  & -17 11 05.5 & 2.2 & 5.51 &\\
PTB17 & 17 57 10.61 & -15 56 20.5 & 3.7 & 3.55 & \\
PTB19 & 17 58 26.35 & -14 25 22.6 & 8.5 & 5.66 & \\
PTB20 & 17 52 14.97 & -11 10 37.6 & 7.6 & 2.05 & \\
PTB23 & 18 31 51.05 & -14 15 19.6 & 10.9 & 10.3 &  \\    
PTB25 & 18 32 04.70 & -13 26 16.3 & 5.9 & 3.86 &  \\
PTB26 & 17 29 12.73 & -16 47 45.7 & 2.4 & 6.20 &  \\
PTB27 & 18 15 13.07 & -23 01 05.2 & 3.2 & 13.5 &\\
PTB30 & 17 51 44.84 & -18 04 25.5 & 3.8 & 32.5 & \\
PTB31 & 18 20 53.71 & -20 48 13.6 & 15.0 & 2.74 & \\
PTB32 & 18 45 11.09 & -23 21 21.7 & 2.9 & 21.7 & \\
PTB34 & 18 30 08.17 & -21 05 10.1 & 3.6 & 10.0 &\\
PTB35 & 18 21 43.32 & -19 39 44.5 & 2.5 & 5.52 &\\
PTB38 & 18 23 04.51 & -17 53 36.2 & 10.1  & 5.52 &\\ 
\hline
\end{tabular}

\medskip{}
\begin{flushleft}

${\rm ^a}$ RA, Dec of the radio source.\\ 
${\rm ^b}$ Difference between the optical and the radio centre.\\
\end{flushleft}
\end{table*}

\section{The 1.3 meter telescope Observations}

\subsection{Imaging}

Optical images of the new PNe candidates were also obtained with the
1.3 m (f/7.7) Ritchey-Cretien telescope at Skinakas Observatory during
2002 in May 19--21, June 10--14 using an \ha $+$\nitrogen\
interference filter (75\AA~bandwidth). The detector was a
1024$\times$1024 SITe CCD with
a field of view of 8.5 $\times$~8.5 arcmin$^{2}$. 
One exposure in the \ha $+$\nitrogen~filter of
1800 s and two exposures in the continuum, each of 180 s, were taken. All new
PNe candidates observed with the \ha $+$\nitrogen\ filter can be seen
in Fig.~\ref{fig02a}. Note that they are at the centre of each image
and an arrow points to their exact position. The image size is
150 arcsec on both sides. North is to the top and east to the left. 
The follow--up observations were performed in the
\ha$+$\nitrogen\ filter to study the morphology of the PNe candidates and 
measure their angular extent.

Following the identification of the PNe candidates, an astrometric
solution in all images was performed using the method presented in
Paper I. Thus, the coordinates of the PNe candidates were calculated
with improved accuracy. The typical rms error in the astrometric
solutions was found to be $\sim$0.3 arcsec. Note that the PN
positional estimates are expected to be accurate to 0.5 to 1.0
arcsec. The coordinates of all new PNe candidates are given in
Table~\ref{table1}.

\subsection{Spectroscopy}

Low dispersion spectra were acquired with the 1.3 m telescope at
Skinakas Observatory during 2002 in June 24--28, July 16--17, 23--25
and August 12, 20--25. A 1300 line mm$^{-1}$~grating was used in
conjunction with a 2000$\times$800 SITe CCD covering wavelengths from
4750\AA\ to 6815\AA. This range was selected in order to observe
simultaneously the \hbeta, the \ha\ and the sulphur lines in a single
spectrum with acceptable resolution. Unfortunately, this hardware
configuration does not allow the coverage of a wider range of
wavelengths as well as the velocity determination.  The slit width was
7.7 arcsec and it was always oriented in the south--north
direction. Note that the $\sim$1\AA/pixel and the dispersion of 1300
line mm$^{-1}$~result in a resolution of $\sim$8 \AA\ and $\sim$11\AA\
in the red and blue wavelengths, respectively. The exposure time of
the individual spectra ranges from 3600 to 3900 s depending on the
observing window of each object.  The spectrophotometric standard
stars HR5501, HR7596, HR9087, HR718, and HR7950 \citep{Ha92} were
observed for the absolute calibration of the spectra. The low
resolution spectra were taken on the relatively bright optical part of
each PN candidate. The actual spectra can be seen in Fig. 3, while in
Table~\ref{table3} the line fluxes, corrected for interstellar
reddening, are given.  The interstellar extinction c(\hbeta) and
observational reddening E$_{\rm B-V}$~were derived using the equations
presented in Paper I. Due to the high interstellar extinction in the
direction of the bulge, a second estimation of E$_{\rm B-V}$ was made
using the SFD code
\citep{Sc98}. The comparison shows a general agreement
between the two E$_{\rm B-V}$~calculations within a 3$\sigma$
error. The signal to noise ratios do not include calibration errors,
which, typically, are less than 10 percent. The absolute \ha\ fluxes
(in units of 10$^{-16}$ erg s$^{-1}$ cm$^{-2}$ arcsec$^{-2}$), the
exposure time of each individual spectra, the interstellar extinction
c(\hbeta) with its estimated error and the reddening E$_{\rm
B-V}$~(resulting from our observations and SFD maps) are listed in
Table~\ref{table4}.

\section{Discussion}

The \ha $+$\nitrogen\ images as well as the low resolution spectra of
the newly discovered PNe candidates were used for a more detailed
study. In Fig. 4(a), the IRAS colour--colour diagram is plotted for 13 of
our objects (including these of Paper I) overlaid on the corresponding
colour--colour plot of known PNe \citep{Ac92b}, for comparison
reasons. The known PNe used in this diagram possess good flux quality
(filled circles) and poor quality (open rectangles). It seems that the
new objects belong, or are close, to the region of PNe according to the
criteria presented in Sect. 2.1. In addition, the diagnostic diagram 
of log(\ha/\nitrogen) vs. log(\ha/\sulfurt) of \cite{G91}, together with 
our measured relative line fluxes confirm the photoionization
origin (Fig. 4(b)) of the candidate PNe. 
The \sulfurt/\ha\ ratios are in all cases less
than 0.3. Nine of them display very low S/N ratio in their \nitrogen\
and \sulfurt\ emission lines and are not presented in Table
\ref{table3}. However, their \sulfurt/\ha\ is less than 0.2.

The angular diameters of the new PNe candidates have been measured in
the \ha$+$\nitrogen\ images acquired with the 1.3 m telescope and were 
established with the method described below. This
method was applied because, especially for the faint PNe, it was
difficult to accurately calculate the diameters because their outer
areas were below the 3$\sigma$ level in the original images. In the
case of elliptical shells, the major and the minor axes are given. The
method used to determine the angular diameters involves the conversion
of their intensity values in a scale ranging from 0 to 1000. In
particular, for intensities higher than 85\% of the maximum, a value
of 1000 was assigned, while for values between 65\% to 85\%, 50\% to
65\%, 35\% to 50\%, 20\% to 35\%, 10\% to 20\% and below 10\% of the
maximum intensity, values of 800, 600, 400, 200, 50, 25 were
given. The advantage of this transformation is that having this wide
range of values, the intensity of the object can be distinguished from
the sky background with much better accuracy, and consequently, their
outer part can be clearly determined. For PNe with well defined outer
edges (i.e. with a steep drop-off of the surface brightness), it was
easy to measure the diameter using either the original
\ha $+$\nitrogen~image or the above method. 
However, in case of faint PNe the direct use of the images results in
larger errors compared to this method. This was also demonstrated by
\cite{Ru04} (and references therein) who showed that the flux drops
very fast between intensity levels in the interval of 15 to 5 per cent
of the peak surface brightness and therefore, it is hard to determine
the actual boundary of the PN.  In our case, values between 10 to 20
per cent of the peak surface brightness contour were selected in order
to measure the diameter of each object and in all cases the estimated
numbers are greater than the 3$\sigma$~limit. An example of the method
is shown in Fig.~\ref{fig05}, where an enlargement of the original \ha
$+$\nitrogen~image of PTB34 can be seen in Fig.5(a). The image of the
same PN is shown in Fig.5(b), which is derived by applying
the method described above.  The angular diameter was measured in both
images but the results from the second image were more accurate. The
resulting images were also used to determine the morphological type of
each PN. All dimensions are given in Table~\ref{table4}. Thirteen of
the new PNe are characterized by diameters less than 20 arcsec 
(Bulge limit -- \citealt{Ga83}), three are close to this limit, and three 
by diameters greater than 30 arcsec. The possibility that the latter
are more evolved and/or are nearby cannot be excluded.

The distribution of the new PNe (a) in galactic latitude and (b) in
angular diameter shows a strong concentration towards the bulge and
especially, when $-6^{o} < b < 6^{o}$ and between 4 to 20 arcsec. Both
results are in agreement with what is expected for PNe in the Galactic
bulge. Note that due to the high interstellar extinction no
observations were performed in the range $-3^{o} < b < 3^{o}$, while,
due to the limited resolution of the survey observations, objects with
diameters smaller than 4 arcsec could not be identified.

A morphological type was assigned to the new PNe (Table~\ref{table4})
according to the classification of \cite{Ma00}. Approximately, 28
percent of the new PNe display spherical, well-defined shells, 42
percent appear elliptical and the rest are bipolar or unclear. The
different morphological PNe classes seen in our survey can be found in
\cite{Bo04}. Objects like PTB34, PTB36 display a ring--like structure,
while for example, PTB26, PTB32 posses incomplete bright
shells. Bright compact nebulae like PTB28, PTB38, PTB42, PTB43 are
also detected. However, their morphological type cannot be resolved
clearly with the current data. Moreover, the morphological class of some PNe
could not be determined unambiguously and even though, these have been
classified as round, the possibility that they are elliptical cannot
be ruled out. The electron densities were found to be different for
each morphological class. The median value for ellipticals is higher
than that for round and bipolar PNe. Also, at a first glance the [N
{\sc ii}]/H$\alpha$~and N/O ratios are higher for round than for
elliptical PNe. The morphology of PNe provides the opportunity to
obtain a better understanding of the evolution of stars. It is
generally accepted that the different PNe morphologies are attributed
to progenitors of different masses. According to \cite{Ph03} (and
references therein), circular PNe arise from stars with low mass
progenitors, the bipolar arise from higher mass stars, while
elliptical originate from a range of masses of the
progenitors. Furthermore, the morphological differences can also be
attributed to the PN stage of evolution. Compact nebulae in their
first stages of expansion appear stellar and relatively bright, while
large nebulae with faint surface brightness are in their late
evolutionary stage.

\begin{table*}
\centering
\caption[]{Line fluxes.}
\label{table3}
\begin{tabular}{llllllll}
\hline
Line (\AA) & PTB26 & PTB27 & PTB28 & PTB29 & PTB30 & PTB31 & PTB32 \\
\hline
\hbeta\ 4861 & 100 (45)$^{\rm a}$ & 100 (21) & 100 (26) & 100 (21) & 100 (14) & 100 (25) & 100 (15) \\
\oxygen\ 4959 & 49 (33) & 330 (42) &  312 (52) & 233 (37) & 252 (24) & 203 (40) & 159 (26) \\
\oxygen\ 5007 & 154 (54) & 926 (78) & 864 (90) & 725 (70) & 770 (43) & 601 (73) & 455 (65) \\ 
\heliumb\ 5411 & $-$ & $-$ & $-$ & $-$ & $-$ & $-$ & $-$ \\
\nitrogena\ 5200 & $-$ & $-$ & 16 (11) & $-$ & $-$ & $-$ & $-$ \\
\nitrogen\ 5755 & $-$ & $-$ & 4 (4) & $-$ & $-$ & $-$ & $-$ \\
\helium\ 5876 & 15 (26) & $-$ & 15 (14) & 11 (10) & $-$ & 14 (15) & 15 (7) \\
\heliumb\ 6234 & $-$ & $-$ & $-$ & $-$ & $-$ & $-$ & $-$\\
\oi\ 6300 & $-$ & 4 (12) & 32 (34) & $-$ & $-$ & $-$ & $-$ \\
\sulfur\ 6312 & $-$ & $-$ & 8 (10) & $-$ & $-$ & $-$ & $-$ \\
\oi\ 6363 & $-$ & $-$ & 8 (11) & $-$ & $-$ & $-$ & $-$ \\
\argonV\ 6435 & $-$ & $-$ & $-$ & $-$ & $-$ & $-$ & $-$ \\
\nitrogen\ 6548 & 34 (50) & 5 (20) & 65 (42) & 20 (14) &  $-$ & 5 (17) & 79 (43)  \\
\ha\ 6563 & 285 (147) & 285 (129) & 285 (98) & 285 (84) & 285 (60) & 285 (177) & 285 (86) \\
\nitrogen\ 6584 & 90 (85) & 12 (28) & 210 (85) & 65 (36)  & 16 (12) & 12 (31) & 236 (81) \\
\helium\ 6678 & 5 (13) & $-$ & 7 (8) & $-$ & $-$ & 4 (15) & 5 (8) \\
\sulfurt\ 6716 & 22 (34) & 4 (20) & 41 (27) & 9 (11) & $-$ & 2 (9) & 36 (27) \\
\sulfurt\ 6731 & 15 (28) & 3 (18) & 44 (30) & 7 (9) & $-$ &  2 (7) & 27 (22) \\
\hline
Line (\AA) &  PTB33 & PTB34 & PTB35 & PTB36 & PTB37 & PTB38 & PTB39  \\
\hline
\hbeta\ 4861 &  100 (11) & 100 (61) & 100 (26) & 100 (13) & 100 (16) & 100 (17) & 100 (35)\\
\oxygen\ 4959 & 200 (23) & 207 (100) & 240 (54) & 303 (33) & 231 (30) & 429 (65) & 389 (77) \\
\oxygen\ 5007 & 637 (44) & 611 (182) & 711 (100) & 919 (63) & 643 (57) & 1242 (125) & 1130 (140) \\ 
\heliumb\ 5411 & $-$ & 8 (25) & $-$ & $-$ & $-$ & $-$ & $-$ \\
\nitrogena\ 5200 & $-$ & $-$ & $-$ & $-$ & $-$ & $-$ & $-$ \\
\nitrogen\ 5755 & $-$ & $-$ & $-$ & $-$ & $-$ & $-$ & $-$ \\
\helium\ 5876 &  9 (5) & 4 (19) & 6 (14) & $-$ & $-$ & 16 (22) & 13 (29) \\
\heliumb\ 6234 & $-$ & $-$ & $-$ & $-$ & $-$ & $-$ & $-$ \\
\oi\ 6300 & $-$ & $-$ & 7 (22) & $-$ & $-$ & 5 (12) & $-$ \\
\sulfur\ 6312 & $-$ & 2 (8) & $-$ & $-$ & $-$ & 2 (5) & $-$ \\
\oi\ 6363 & $-$ & $-$ & 3 (16) & $-$ & $-$ & 2 (3) & $-$ \\
\argonV\ 6435 & $-$ & 1 (7) & $-$ & $-$ & $-$ & $-$ & $-$ \\
\nitrogen\ 6548 & 38 (16) & 0.4 (8) & 184 (142) & $-$ & $-$ & 9 (32) & 2 (11) \\
\ha\ 6563 & 285 (47) & 285 (250) & 285 (195) & 285 (76) & 285 (85) & 285 (225) & 285 (190) \\
\nitrogen\ 6584 &  105 (28) & 2 (8) & 569 (255) & $-$ & $-$ & 29 (72) & 3 (16) \\
\helium\ 6678 & $-$ & 2 (15) & 5 (19) & $-$ & $-$ & 3 (16) & 2 (14) \\
\sulfurt\ 6716 & 30 (12) & 0.8 (9) & 20 (41) & $-$ & $-$ & 3 (13) & $-$ \\
\sulfurt\ 6731 & 27 (10) & 0.6 (8) & 15 (33) & $-$ & $-$ & 6 (23) & $-$ \\
\hline
Line (\AA) & PTB40 & PTB41 & PTB42$^{\rm a}$ & PTB43$^{\rm a}$ & PTB44 \\
\hline
\hbeta\ 4861 &  100 (25) & 100 (24) & 100 (24) & 100 (45) & 100 (35) \\
\oxygen\ 4959 & 376 (57) & 212 (43) & 895 (89) & 238 (80) & 363 (51)\\
\oxygen\ 5007 & 1146 (103) & 609 (79) & 2620 (161) &  710 (147) & 1085 (96)\\ 
\heliumb\ 5411 & $-$ & 9 (8) & 5 (6) & $-$ & $-$ \\
\nitrogena\ 5200 & $-$ & $-$ & $-$ & $-$ & $-$ \\ 
\nitrogen\ 5755 & $-$ & $-$ & 44 (32) & 3 (13) & $-$ \\
\helium\ 5876 & 4 (8) & $-$ & 13 (12) & 19 (44) & $-$\\
\heliumb\ 6234 & $-$ & $-$ & $-$ & $-$ & $-$ \\
\oi\ 6300 & $-$ & $-$ & 20 (21) & 3 (18) & $-$ \\
\sulfur\ 6312 & $-$ & $-$ & $-$ & 5 (23) & $-$ \\
\oi\ 6363 & $-$ & $-$ & $-$ & $-$ & $-$ \\
\argonV\ 6435 & $-$ & $-$ & $-$ & $-$ & $-$ \\
\nitrogen\ 6548 & $-$ & $-$ & 182 (83) & 5 (32) & $-$\\
\ha\ 6563 & 285 (101) & 285 (140) & 285 (99) & 285 (219) & 285 (89)\\
\nitrogen\ 6584 & $-$ & 2 (4) & 607 (147) & 16 (65) & 10 (8) \\
\helium\ 6678 & $-$ & $-$ & 4 (8) & 4 (25) & $-$\\
\sulfurt\ 6716 & $-$ & $-$ & $-$ & 1 (11) & $-$ \\
\sulfurt\ 6731 & $-$ & $-$ & $-$ & 2 (16) & $-$ \\
\hline
\end{tabular}

\medskip{}
\begin{flushleft}

${\rm ^a}$ Numbers in parentheses represent the signal to noise ratio of the line fluxes, measured at the center of the corresponding emission line profile.\\ 
${\rm }$ All fluxes are normalized to F(\hbeta)=100 and are corrected for interstellar extinction.\\
\end{flushleft}
\end{table*}

\begin{table*}
\centering
\caption{Exposure time and basic physical parameters.}
\label{table4}
\begin{tabular}{c|c|c|c|c|c|c|c|c|c|c|c}
\hline
Object & Exp. time$^{\rm 1}$ & Diameter$^{\rm 2}$ &M.Type$^{\rm 3}$ & F(\ha)$^{\rm 4}$ & c(\hbeta)$^{\rm 5}$ & $\sigma_{\rm c(H\beta)}^{\rm 6}$ & E$_{\rm B-V (OBS)}^{\rm 7}$ & $\sigma_{\rm E_{B-V}}^{\rm 8}$  & E$_{\rm B-V (SFD)}^{\rm 9}$  &n$_{\rm [S~{\sc II}]}^{\rm 10}$ &  $\sigma_{\rm n_{\rm [S~{\sc II}]}}^{\rm 11}$\\
\hline
PTB26 & 3600 & 28.5$\times$16.0 & B & 5.8 & 0.62 & 0.03 & 0.43 & 0.02 & 0.39 & $<$0.07 & $-$ \\
PTB27 & 3600 & 12.0 & R & 4.7 & 1.86 & 0.10 & 1.29 & 0.07 & 1.45 & 0.36 & 0.15 \\
PTB28 & 3600 & 4.5 & $-$ & 4.1 & 0.69 & 0.05 & 0.48 & 0.03 & 0.48 & 0.65 & 0.03 \\
PTB29 & 3600 & 37.0$\times$34.0 & E & 7.4 & 0.92 & 0.04 & 0.64 & 0.03 & 0.66 & 0.16 & 0.03 \\
PTB30 & 3600 & 26.0 & R & 2.1 & 1.57 & 0.15 & 1.08 & 0.10 & 0.99 & $-$ & $-$ \\
PTB31 & 3900 & 15.5 & R & 10.7 & 2.72 & 0.14 & 1.88 & 0.10 & 1.91 & 0.67 & 0.33 \\
PTB32 & 3600 & 135.0$\times$115.0 & E & 1.5 & 0.74 & 0.11 & 0.51 & 0.08 & 0.39 & $<$0.07 & $-$ \\
PTB33 & 3600 & 17.0 & R & 2.0 & 0.58 & 0.11 & 0.39 & 0.08 & 0.49 & 0.38 & 0.07 \\
PTB34 & 3900 & 12.5 & R & 36.6 & 1.08 & 0.02 & 0.74 & 0.01 & 0.71 & $<$0.07 & $-$ \\
PTB35 & 3900 & 15.0$\times$11.5 & E & 16.6 & 1.89 & 0.05 & 1.31 & 0.03 & 1.47 & $<$0.07 & $-$ \\
PTB36 & 3900 & 22.0 & B & 2.5 & 1.02 & 0.16 & 0.71 & 0.11 & 0.62 & $-$ & $-$ \\
PTB37 & 3900 & 16.0$\times$14.0 & E & 4.3 & 1.01 & 0.07 & 0.71 & 0.05 & 0.68 & $-$ & $-$ \\
PTB38 & 3900 & 5.5 & $-$ & 20.7 & 2.47 & 0.10 & 1.71 & 0.07 & 2.07 & 27.0 & 4.7 \\
PTB39 & 3900 & 8.0$\times$6.5 & E & 12.0 & 1.46 & 0.03 & 1.01 & 0.03 & 1.01 & $-$ & $-$ \\
PTB40 & 3900 & 11.5$\times$10.0 & E & 10.9 & 0.99 & 0.06 & 0.69 & 0.04 & 0.46 & $-$ & $-$ \\
PTB41 & 3600 & 10.5$\times$8.0 & E & 11.2 & 1.72 & 0.06 & 1.19 & 0.03 & 1.08 & $-$ & $-$ \\
PTB42 & 3600 & 4.0 & $-$ & 12.7 & 1.41 & 0.04 & 0.98 & 0.03 & 0.69 & $-$ & $-$ \\
PTB43 & 3900 & 5.0 & $-$ & 26.6 & 1.24 & 0.02 & 0.86 & 0.01 & 0.62 & 4.4 & 1.09 \\
PTB44 & 3900 & 58.0 & E & 4.5 & 0.92 & 0.04 & 0.64 & 0.03 & 0.24 & $-$ & $-$ \\
\hline
\end{tabular}

\medskip{}
\begin{flushleft}

$^{\rm 1}$ Total exposure time in sec. \\

$^{\rm 2}$ Optical diameters in arcsec. \\ 

$^{\rm 3}$ Morphological classification according to Manchado et al. (2000), where R($=$round), E ($=$elliptical) and B ($=$bipolar).\\ 

$^{\rm 4}$ Absolute \ha\ flux in units of \flux. \\

$^{\rm 5}$ Logarithmic extinction at \hbeta\ (see Sect. 3.2). \\

$^{\rm 6}$ 1$\sigma$~error on the logarithmic extinction. \\ 

$^{\rm 7}$ Observed reddening (see Sect. 3.2). \\

$^{\rm 8}$ Error on the observed reddening.\\

$^{\rm 9}$ Reddening according to the maps of Schlegel et al. (1998) (see Sect. 3.2). \\

$^{\rm 10}$ Electron density in units of 10$^{3}$~cm$^{-3}$ (see Sect. 5).\\ 

$^{\rm 11}$ Error on the electron density.\\

\end{flushleft}
\end{table*}

Following the procedure described in Paper I and assuming a
temperature of 10$^{4}~$K \citep{Cu00}, an estimation of the electron
density n$_{\rm [S~{\sc II}]}$ was made, with the ``temden'' task of
the ``nebular'' package in IRAF, for each specific Sulfur line ratio
(\sulfurt\ 6716\AA/\sulfurt\ 6731\AA). The electron densities and
their associated errors are given in Table~\ref{table4}. In addition,
using the observed \nitrogen\ 6548$+$6584\AA/\nitrogen\ 5755\AA\
ratios, an estimate of the electron temperature T$_{\rm [N~{\sc
II}]}$~was made, whenever possible. In fact, only for three of our new
PNe temperatures close to 10000 K were found. The high interstellar
extinction towards the Galactic bulge as well as the low surface
brightness for many of the new PNe prevented the registration of high
quality spectra. This affects mainly the lower intensity lines like
\sulfurt\ or
\heliumb~5411~\&~6234~\AA, which are faint or even undetectable. For
all the PNe candidates in our survey, \ha, \oxygen\ and \hbeta\
emission lines could be measured and in most cases, \NII. However, the
spectra of some of the new PNe (e.g. PTB36) have a low signal to noise
and only \hbeta, \ha\ and \oxygen\ are well determined. Since these
PNe are found in a region of significant extinction shorter
wavelengths are more affected than longer wavelengths.

In Table 3 we list the fluxes, measured in our long--slit spectra,
corrected for interstellar extinction, where it is evident that almost
all PNe display very strong \oiii\ emission relative to \ha. It is
known that the N/O and He abundances reflect the mass distribution of
the progenitor star, with more massive objects having higher N/O and
He abundances in comparison with the low--mass objects
\citep{Cu00}. In our case, the abundances in N/O and He are generally
low implying old progenitor stars. The latter suggests that they
belong to the Galactic bulge region according to \cite{We88} and
\cite{Cu00}. Both authors showed that the Galactic bulge PNe originate
from old progenitors with low N/O and He abundances. The excitation
classes of our new PNe were studied according to the classification
criteria of \cite{Al56}, \cite{Fe68} and \cite{We75}. They are derived
basically from the ratios of \oxygen/\hbeta, \heliumb\ 4686\AA/\hbeta,
\oii\ 3727 \AA/\oxygen\ 4959 \AA\ and \heliumb\ 4686 \AA/\helium\ 5876
\AA. In our case, following the relation 2.1a of \cite{Do90}, the
\oxygen/\hbeta\ ratio suggests that our new PNe belong to the low
(2--4) and medium (4--6) excitation classes. It is only PTB42 showing
an excitation class greater than 6 and therefore, it must be highly
ionized. However, the \heliumb\ 4686 \AA\ and \oii\ 3727 \AA\ emission
lines are outside the observed range due to the hardware configuration
(see Sect. 3.2) and thus, it is not possible to better confine the
actual excitation class of each PN.

\section{Conclusions}

We presented results from the narrow band \oiii\ survey of PNe in the
Galactic bulge ($l>0^{o}$). Covering the remaining 34 percent of our
selected region, we detected 60 objects, including 41 known PNe and 19
new PNe. \ha $+$\nitrogen\ images as well as low resolution spectra of
the new PNe were taken which confirmed the photoionized nature of the
emission. About 84 percent of the detected objects have angular sizes
$\leq$ 20--25 arcsec, while all show \ha/\oiii\ ratios less than 1. 
Four (4) of our new PNe are found to be
associated with IRAS sources. Radio sources for fifteen (15) of our
new PNe (including those of paper I) were identified. All new PNe display
low N/O and He abundances implying old progenitor stars, which is one
of the characteristics of the Galactic bulge PNe.  
In a forthcoming
paper (paper III), the use of the photoionization model CLOYDY will
provide a deeper insight to the physical parameters of the new
PNe. The information presented in papers I and II, along with the
results that will be obtained from the CLOUDY model will offer a
valuable tool to studies of the dynamics and kinematics of the
Galactic bulge.

\begin{figure*}
\centering
\scalebox{0.90}{\includegraphics{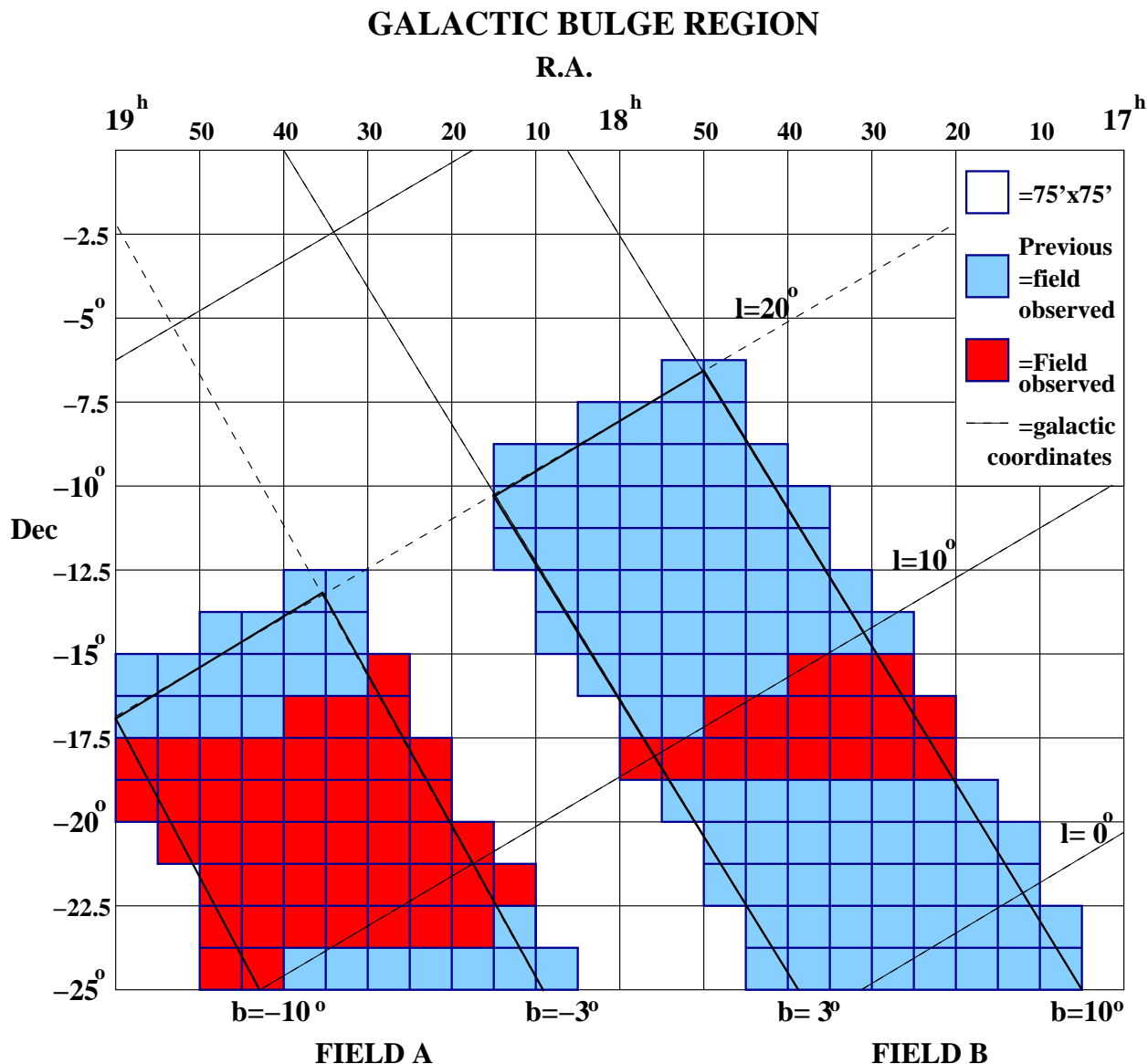}}
\caption[]{Optical imaging survey grid in equatorial
coordinates. Galactic coordinates are also included (dash lines) to
permit an accurate drawing of the selected Bulge region (bold solid
lines). The dark filled and light filled rectangles represent the
remaining observed field and the observed fields in the year 2000,
respectively.}
\label{fig01}
\end{figure*}

\begin{figure*}
\centering
\scalebox{0.90}{\includegraphics{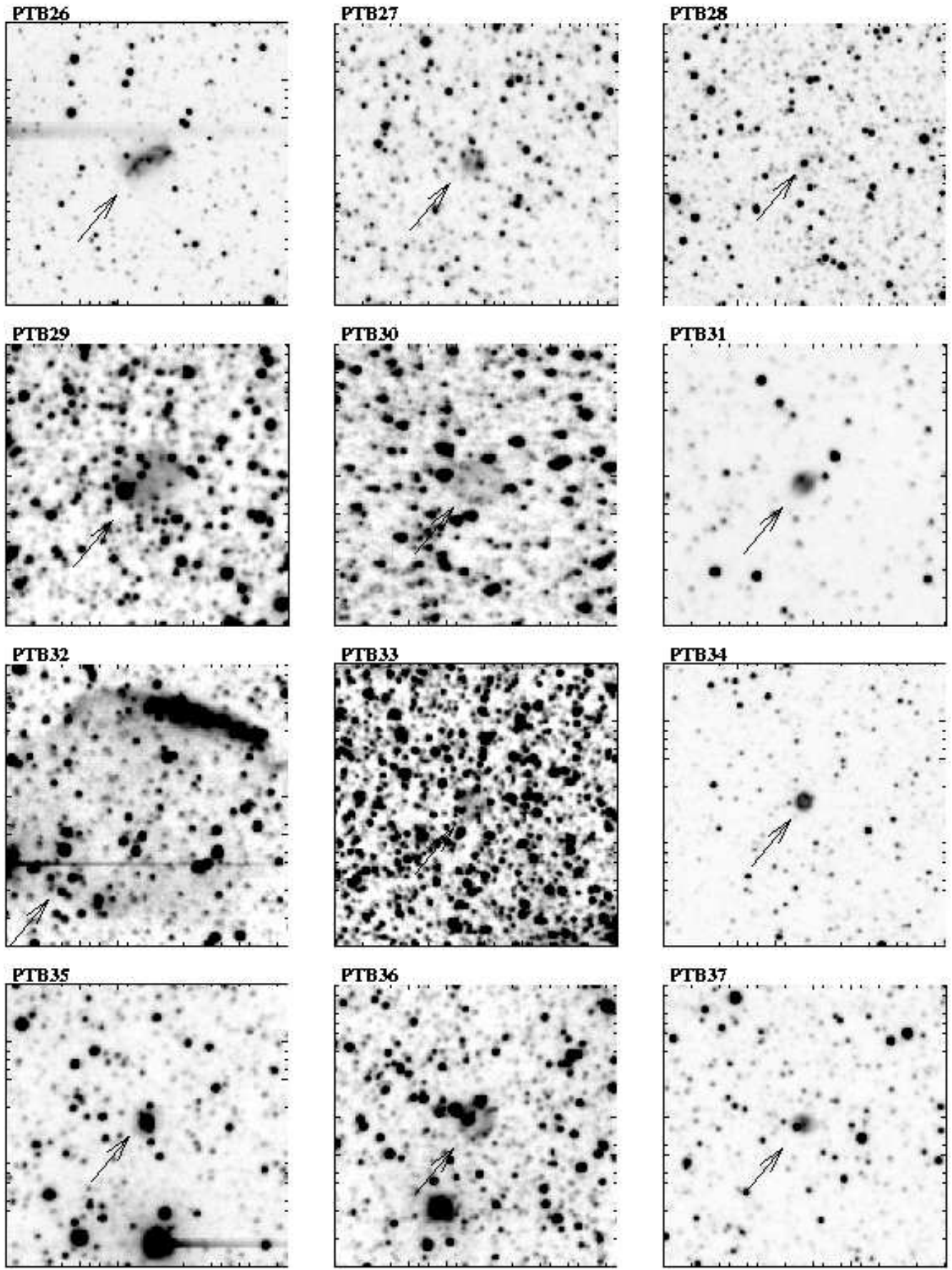}}
\caption[]{\ha $+$\nitrogen~images of all new PNe taken with the 1.3 m
telescope. The arrows point to their position. The images have a size
of 150\arcsec on both sides. North is at the top, East to the left.}
\label{fig02a}
\end{figure*}

\begin{figure*}
\centering
\addtocounter{figure}{-1}
\scalebox{0.90}{\includegraphics{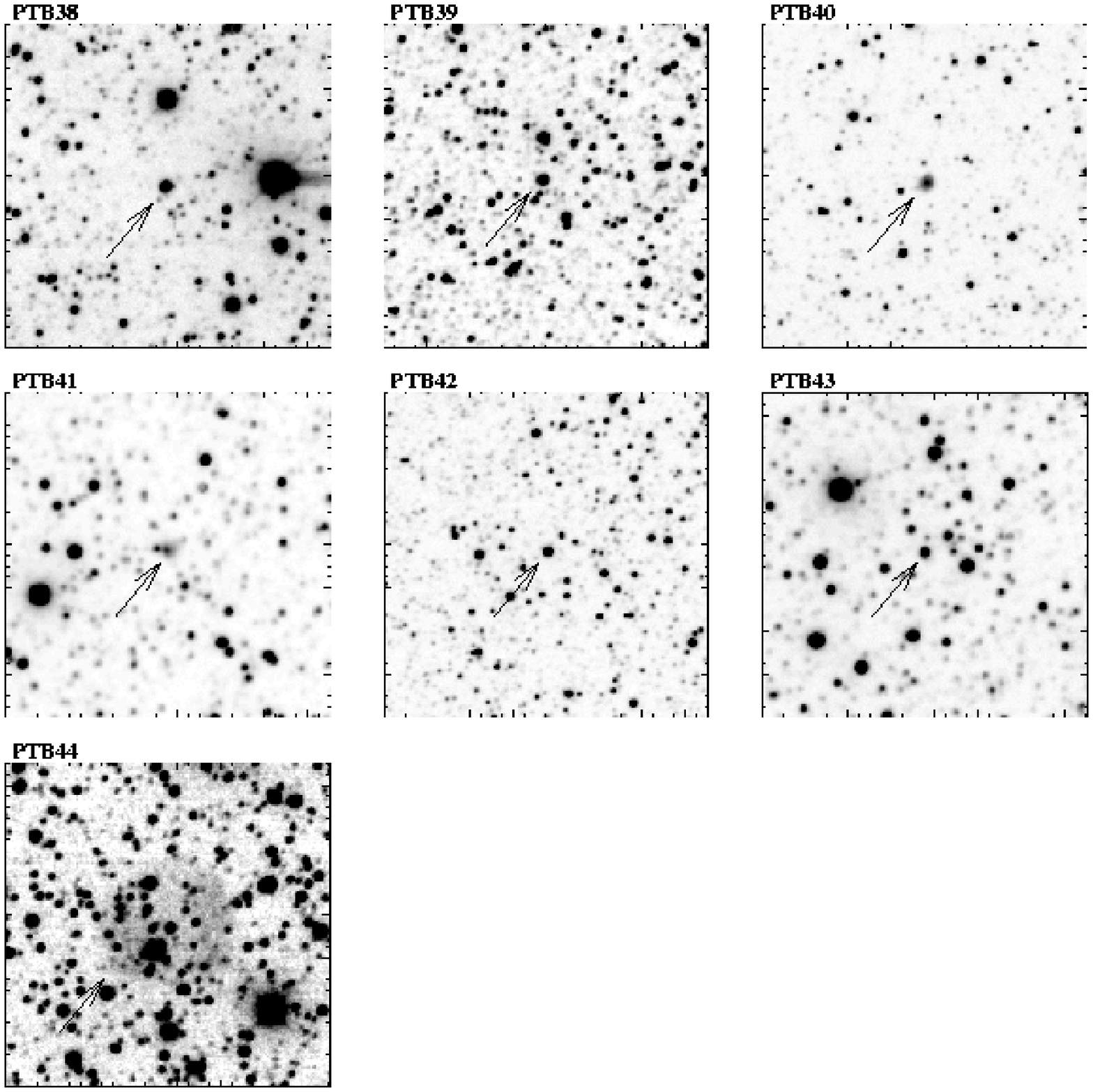}}
\caption[]{continued} 
\label{fig02b}
\end{figure*}

\begin{figure*}
\centering
\scalebox{0.90}{\includegraphics{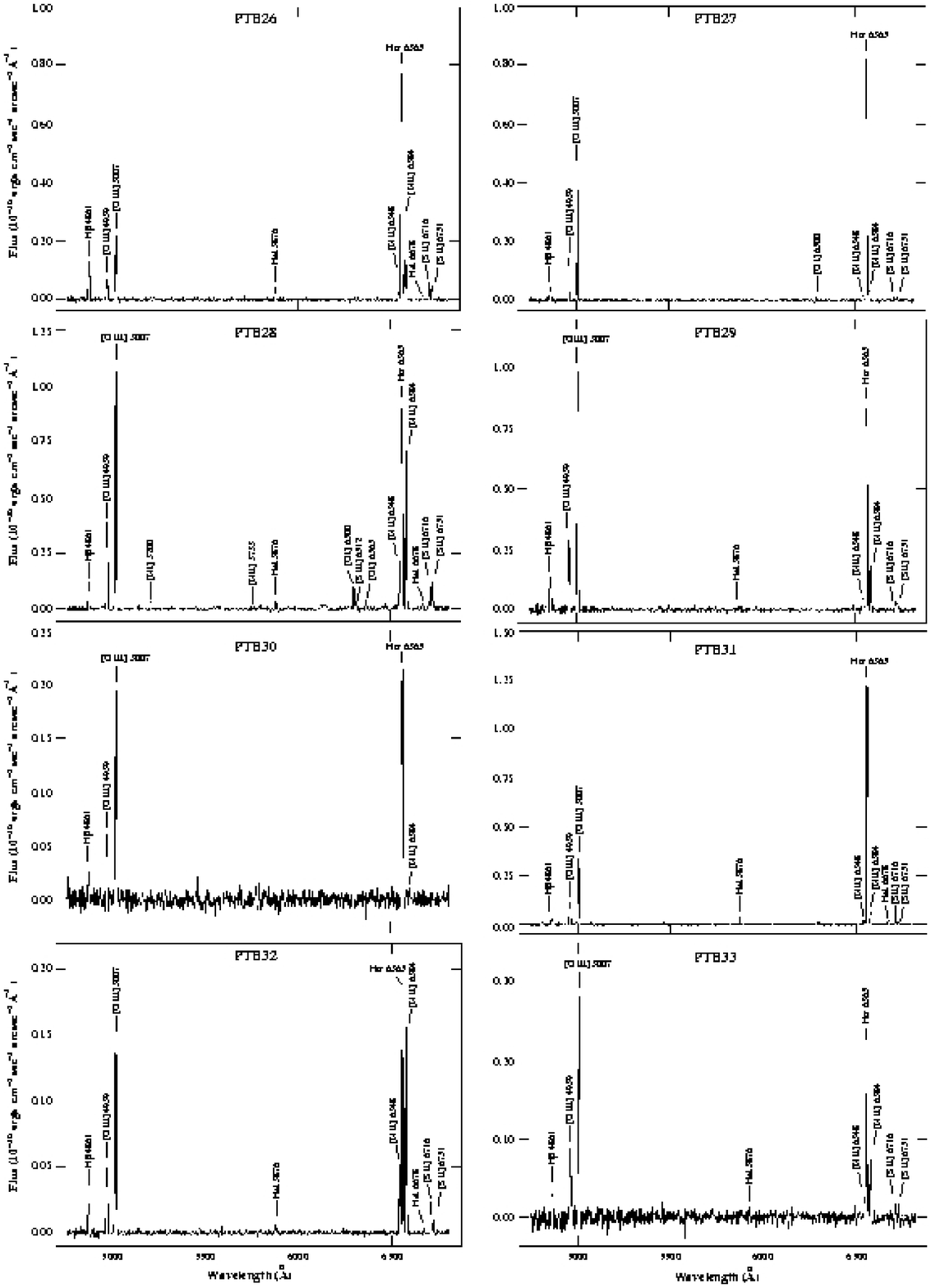}}
\caption[]{Observed spectra of our new PNe taken with the 1.3
m telescope. They cover the range of 4750\AA\ to 6815\AA\ and the
emission line fluxes (in units of $10^{-16}$ erg s$^{-1}$ cm$^{-2}$
arcsec$^{-2}$ \AA$^{-1}$) are corrected for atmospheric
extinction. Line fluxes corrected for interstellar extinction are
given in Table 2.}
\label{fig03a}
\end{figure*}

\begin{figure*}
\centering
\addtocounter{figure}{-1}
\scalebox{0.90}{\includegraphics{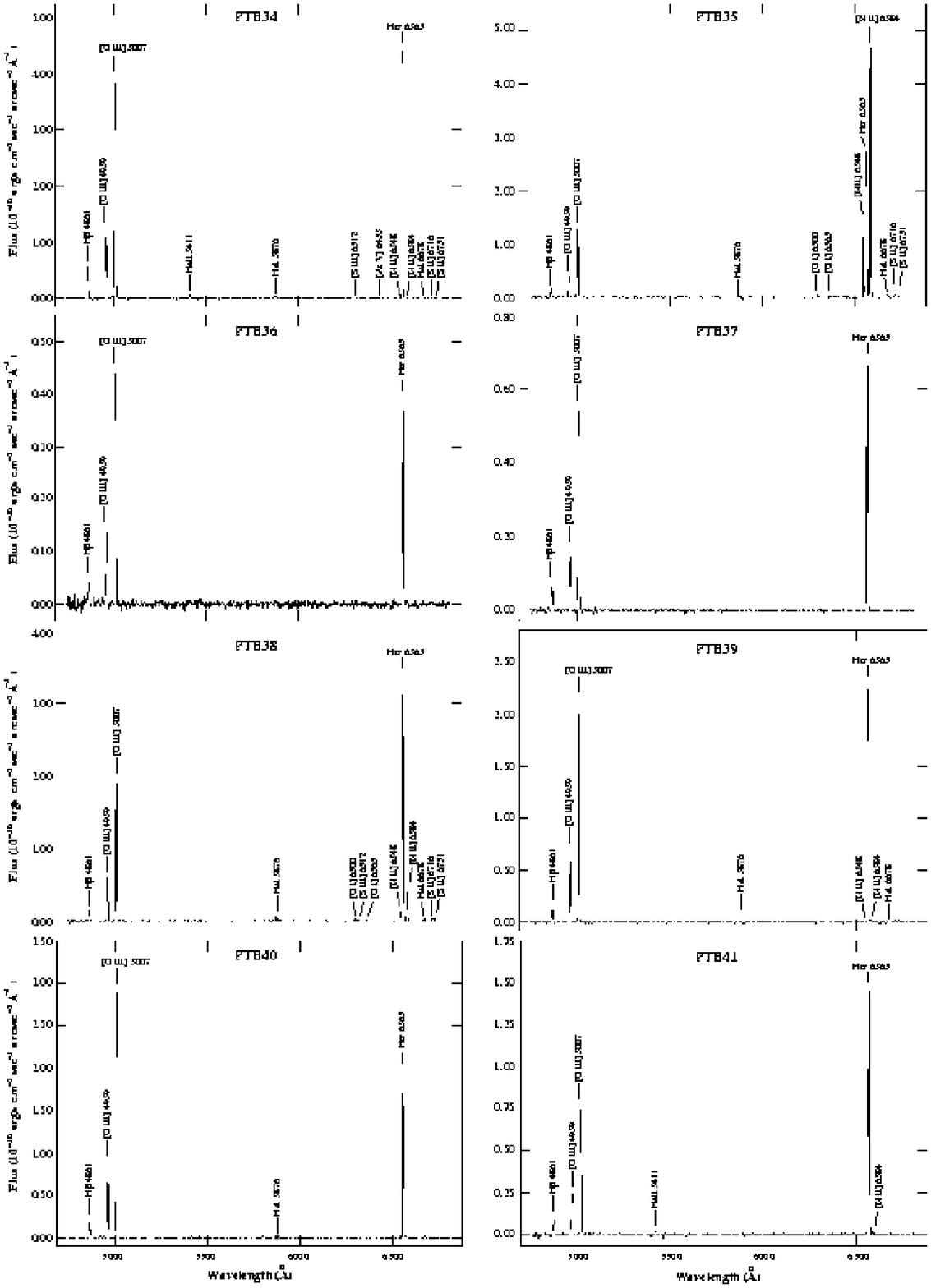}}
\caption[]{continued}
\label{fig03b}
\end{figure*}

\begin{figure*}
\centering
\addtocounter{figure}{-1}
\scalebox{0.65}{\includegraphics{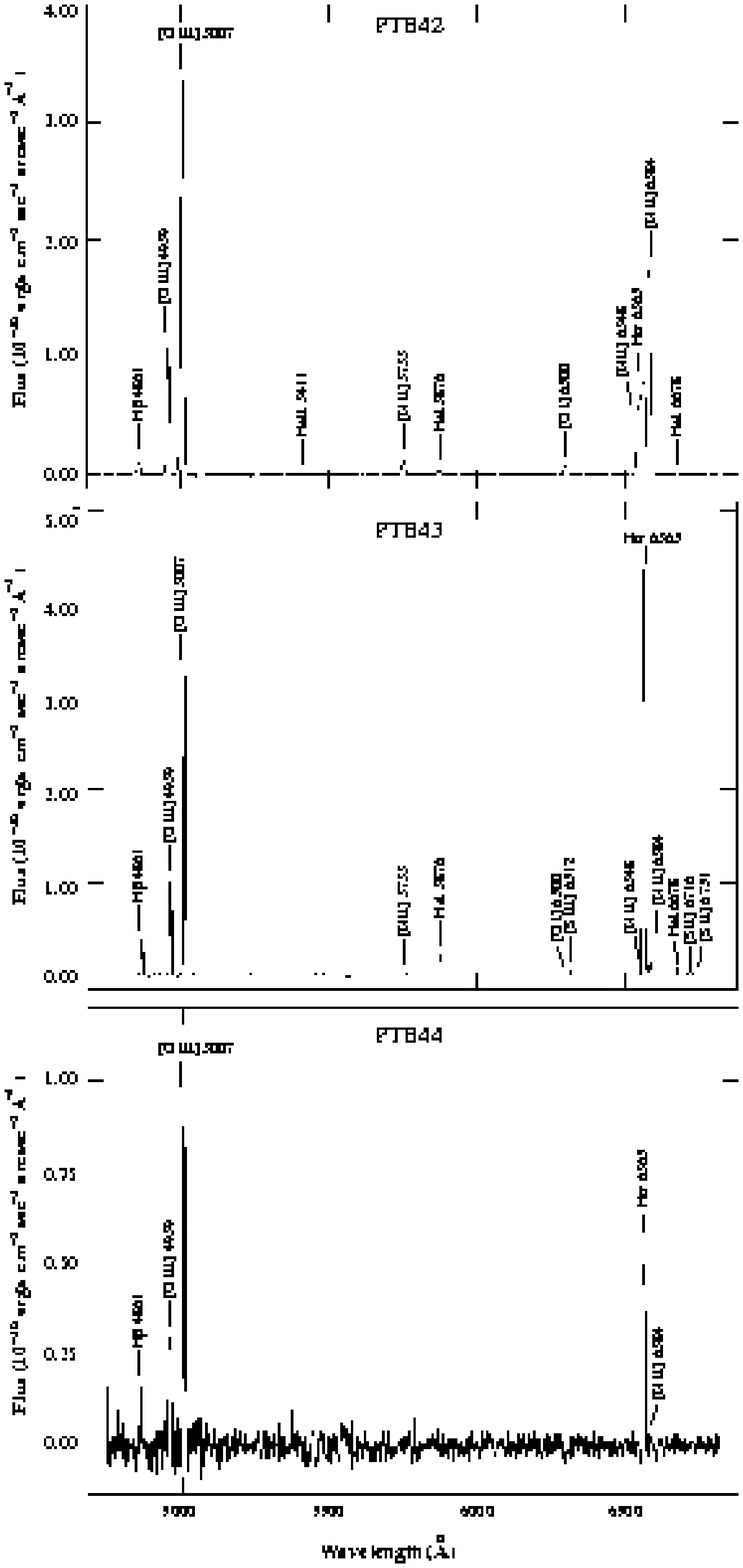}}
\caption[]{continued}
\label{fig03c}
\end{figure*}

\begin{figure*}
\centering
\scalebox{0.65}{\includegraphics{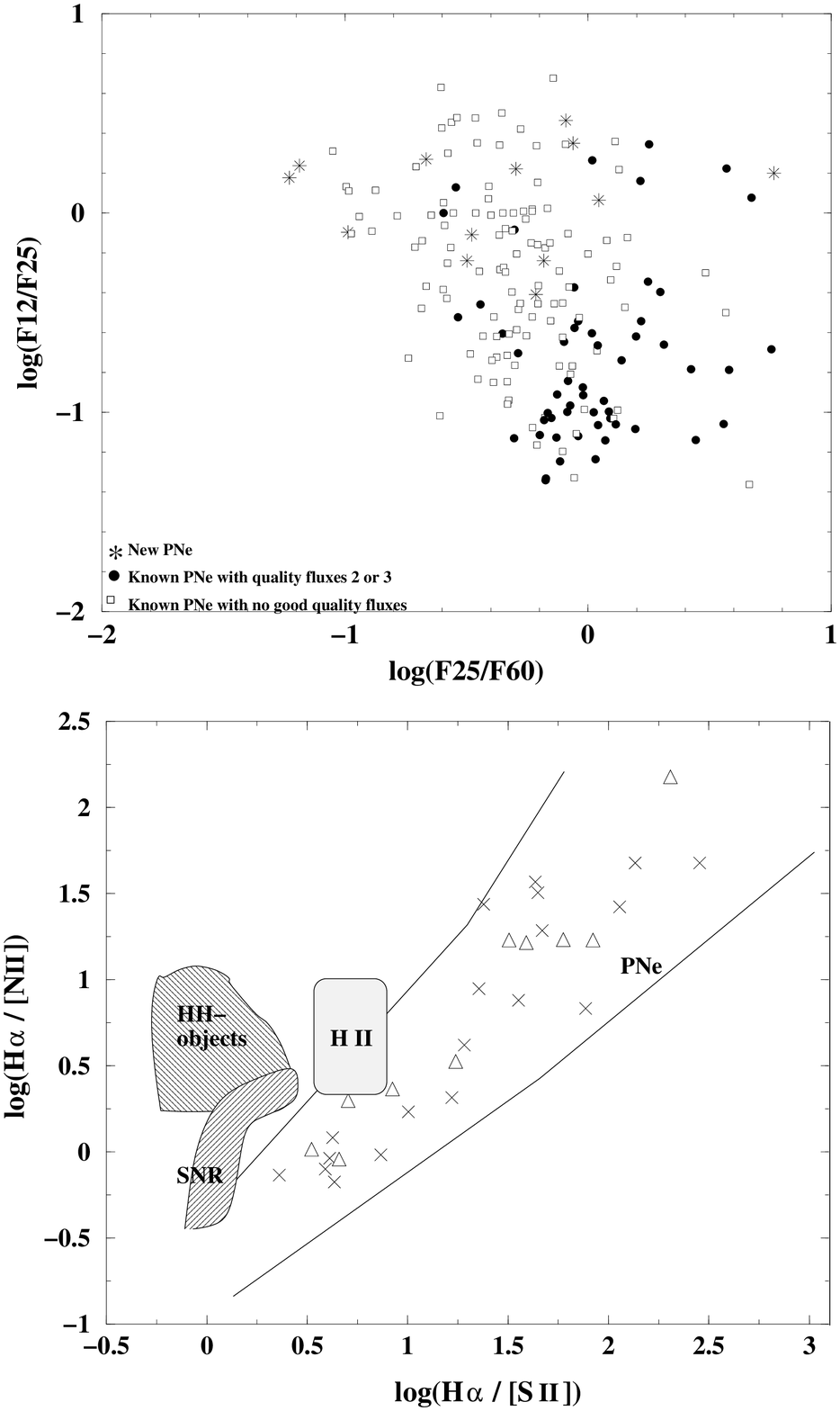}}
\caption[]{(a) IRAS colour--colour diagram of 13 of our objects
(stars), overlaid on the colour--colour diagram of Acker's catalogued
PNe with good (circles) and not good (rectangles) quality fluxes and
(b) Diagnostic diagram (Garcia et al. 1991), where the positions of
the new PNe are shown with a triangle ($\bigtriangleup$). For
comparison, the position of (i) the new PNe presented in paper I are
also shown with a cross (X) and (ii) other objects (supernova remnants
- SNR, H {\sc ii} regions and Herbig Haro objects - HH) are shown,
too.}
\label{fig04}
\end{figure*}

\begin{figure*}
\centering
\scalebox{0.80}{\includegraphics{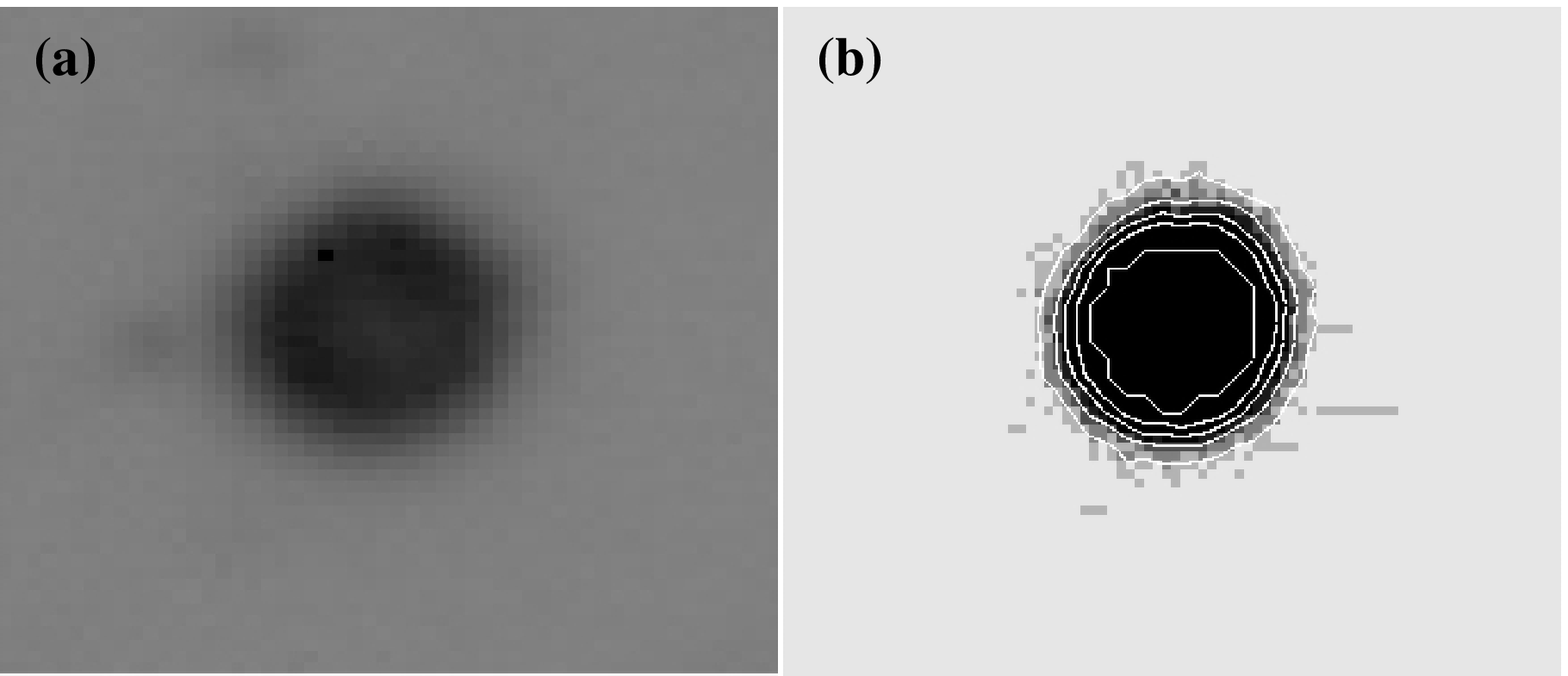}}
\caption[]{(a) Enlargement of the \ha $+$\nitrogen~image of PTB34 in
high contrast in order to show its outer halo (b) Image of the same PN
overlaid with the different scale contours as derived from the method
described in the text. North is in the top, east to the left in both figures.}
\label{fig05}
\end{figure*}

\section*{Acknowledgments} 

The authors would like to thank the referee Q. Parker for his
comments and suggestions and for providing us  the
Edinburgh/AAO/Strasbourg Catalogue of Galactic Planetary Nebulae 
and the Macquarie/AAO/Strasbourg \ha\ PN Catalogue.
We would also like to thank A. Acker for providing us the First Supplement 
to the Strasbourg--ESO Catalogue of Galactic Planetary Nebulae.
The help of E. Semkov, G. Paterakis, A. Kougentakis and A. Strigachev 
during these observations is acknowledged. 
Skinakas Observatory is a collaborative project of
the University of Crete, the Foundation for Research and
Technology-Hellas, and the Max-Planck-Institut f\"{u}r
extraterrestrische Physik.

\label{lastpage}
\end{document}